\documentclass{grant_nature}%twocolumn,showpacs,showkeys,amsmath,amssymb,prl]{revtex4-1}
\usepackage{graphicx}
\usepackage{amssymb}
\usepackage{amsmath}
\usepackage{subfigure}
\usepackage{color}
\usepackage{xcolor}
\usepackage{float}
\usepackage[switch, modulo]{lineno}
\setpagewiselinenumbers
\modulolinenumbers[5]
\linenumbers
\usepackage{rotating}
\usepackage{multirow}

\usepackage[symbol]{footmisc}

\begin{document}
\nolinenumbers

\title{Experiments conducted in the burning plasma regime with inertial fusion implosions} 
\author{
J. S. Ross$^{1,*}$,
J. E. Ralph$^{1,*}$,
A. B. Zylstra$^{1,*}$,
A. L. Kritcher$^{1}$,
H. F. Robey$^{2}$,
C. V. Young$^{1}$,
O. A. Hurricane$^{1}$,
D. A. Callahan$^{1}$,
K. L. Baker$^{1}$,
D. T. Casey$^{1}$
% ...
\\ 
\\ $^1$ Lawrence Livermore National Laboratory, P.O. Box 808, Livermore, California 94551-0808, USA
\\ $^2$ Los Alamos National Laboratory, Mail Stop F663, Los Alamos, New Mexico 87545, USA
\\ $^3$ Laboratoire pour l'utilisation des Lasers Intenses chez {\'E}cole Polytechnique, F-91128 Palaiseau cedex, France
\\ $^4$ Laboratory for Laser Energetics, University of Rochester, Rochester, New York 14623, USA
\\ $^5$ General Atomics, San Diego, California 92186, USA
\\ $^6$ Massachusetts Institute of Technology, Cambridge, Massachusetts 02139, USA
\\ $^7$ SLAC National Accelerator Laboratory, Menlo Park, California 94025, USA
\\ $^8$ Atomic Weapons Establishment, Aldermaston, RG7 4PR, United Kingdom
\\ $^9$ Diamond Materials Gmbh, 79108 Freiburg, Germany
\\ $^{*}$ These authors contributed equally: J.S. Ross, J.E. Ralph, and A.B. Zylstra}
%\\ $^3$ General Atomics, P.O. Box 85608, San Diego, California 92186-5608  

\date{\today}

\maketitle

\begin{abstract}
An experimental program is currently underway at the National Ignition Facility (NIF) to compress deuterium and tritium (DT) fuel to densities and temperatures sufficient to achieve fusion and energy gain. The primary approach being investigated is indirect-drive inertial confinement fusion (ICF), where a high-Z radiation cavity (a hohlraum) is heated by lasers, converting the incident energy into x-ray radiation which in turn drives the DT fuel filled capsule causing it to implode. Previous experiments reported DT fuel gain exceeding unity [O.A. Hurricane et al., Nature 506, 343 (2014)] and then exceeding the kinetic energy of the imploding fuel [S. Le Pape et al., Phys. Rev. Lett. 120, 245003 (2018)]. We report on recent experiments that have achieved record fusion neutron yields on NIF, greater than 100 kJ with momentary fusion powers exceeding 1PW, and have for the first time entered the burning plasma regime where fusion alpha-heating of the fuel exceeds the energy delivered to the fuel via compression. This was accomplished by increasing the size of the high-density carbon (HDC) capsule, increasing energy coupling, while controlling symmetry and implosion design parameters.  Two tactics were successful in controlling the radiation flux symmetry and therefore the implosion symmetry: transferring energy between laser cones via plasma waves, and changing the shape of the hohlraum.  In conducting these experiments, we controlled for known sources of degradation.  Herein we show how these experiments were performed to produce record performance, and demonstrate the data fidelity leading us to conclude that these shots have entered the burning plasma regime.
\end{abstract}

\begin{figure*}[t]
	\includegraphics[width=7.0 in]{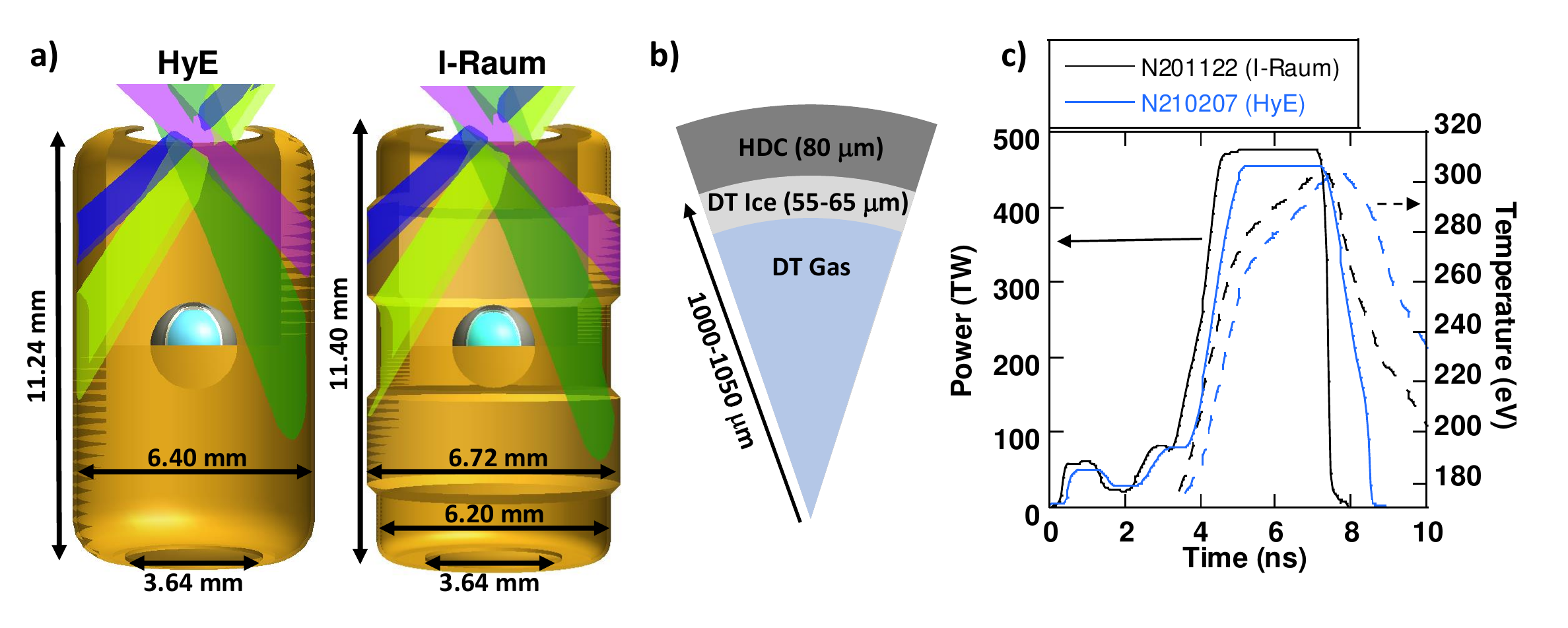}
	\caption{\label{setup}a) The experimental setup is shown for the HyE (left) cylinder and I-Raum (right). The NIF lasers are grouped into four cones at angles relative to the axis of 23$^\circ$ and 30$^\circ$ (the inners, one beam from each cone is shown in shades of green), and the outer cones at 44$^\circ$ and 50$^\circ$ (purple and blue, respectively). Half of the beams enter the hohlraum through a laser entrance hole at the top of the target and half through the bottom. b)  Wedge schematic of the capsule, showing the features versus radius.  c) Two representative laser beam pulse shapes (solid lines) for the HyE and I-Raum experiments are shown and compared to the measured radiation temperatures (dashed lines).}
	\label{fig:setup}
\end{figure*}

The experiments were performed at the National Ignition Facility (NIF) \cite{nif} where 192 laser beams deliver up to 1.9 MJ of 351 nanometer laser light to the inside of a hohlraum (Fig. \ref{setup}), creating an x-ray radiation bath with a maximum radiation temperature ($T_r$) of $\sim$300 eV (nearly 3.5 million degrees C).  The x-ray drive is absorbed by a spherical capsule, filled with cryogenic DT fuel, and mounted in the center of the hohlraum. The capsule is comprised of an outer ablator layer surrounding the inner layer of DT; x rays are absorbed at the outer surface of the ablator, triggering the inward compression of the fuel. The energy coupled to the capsule,
\begin{equation} 
	E_{cap} \propto \int T_{r}(t)^4 r_{cap}(t)^2 dt,
\end{equation}
is a function of $T_r$, the time-dependent outer capsule radius ($r_{cap}$), and the implosion duration, which scales as the initial radius. Our experimental campaigns were designed around a strategy\cite{ISI:000451447200001} to increase the capsule radius by $\sim 10-15 \%$ relative to previous high-performing experiments, which predominantly used high-density carbon (HDC) ablators\cite{doi:10.1063/1.5033459, doi:10.1063/1.4982215,PhysRevLett.120.245003,Baker2018,Baker2020} with inner radii between 910-950 $\mu$m. Our goal was therefore to use 1000 and 1050 $\mu$m inner radius capsules with comparable $T_r$ and other implosion parameters constant\cite{KritcherYoungRobeyBurningPlasmaNatPhys} to these previous works, resulting in an increased $E_{cap}$ of $\sim 30-40\%$. Maintaining $T_r$ necessitates a similar-sized hohlraum; increasing the capsule size relative to the hohlraum is beneficial for efficiency but challenging for low-mode symmetry control, especially the polar mode 2 (e.g. the amplitude of a Legendre polynomial, $P_2$).  Without additional control over $P_2$ these designs would have produced highly-oblate implosions, well outside of the specification\cite{Kritcher2014} of $\pm 4$ $\mu$m. Multiple tactics for maintaining symmetry control have been investigated\cite{doi:10.1063/1.4982215}; two proved successful. By controlling low-mode symmetry and creating high-quality larger-scale implosions we have produced implosions that have entered the burning plasma regime; analysis of the burning-plasma criteria is presented in a companion paper by Zylstra, Hurricane et al. (Ref. \citenum{ZylstraBurning}), computational design and analysis is presented in a companion paper by Kritcher, Young, Robey et al. (Ref. \citenum{KritcherYoungRobeyBurningPlasmaNatPhys}). 
%are described briefly below and in detail in Ref. \citenum{KritcherYoungRobeyBurningPlasmaNatPhys}. 

Two approaches were used to control the $P_2$ symmetry. Using `wavelength detuning' ($\Delta \lambda$) has been explored in the `Hybrid E' (HyE) campaign\cite{Zylstra2021}. As the laser beams enter the hohlraum, a process known as crossed-beam energy transfer (CBET)\cite{michel_tuning_2009} exchanges energy between the outer and inner cones (see Fig. \ref{setup}); the NIF beams are grouped into inner cone beams with 23 and 30 angle of incidence relative to the hohlraum axis and outer cone beams with 44 and 50 angle of incidence). CBET transfers energy between the beams via stimulated Brillouin scattering and depends on the local plasma conditions and wavelength separation between beams. This tactic for controlling symmetry is to introduce a small $\Delta \lambda$) between the cones, which alters the amount of energy transferred\cite{michel_tuning_2009}. The $\Delta \lambda$ is fixed in time and transfer throughout the pulse is compensated using simulations to tune the pulse shape to achieve the desired time-dependent flux symmetry\cite{KritcherYoungRobeyBurningPlasmaNatPhys}\footnote[1]{This tactic was previously used in hohlraums with a high gas-fill density, which required high values of $\Delta \lambda$ ($>$ 7$\AA$). This led to difficulties controlling the time-dependent symmetry, and diminishing returns due to absorption of the inner beams in the gas fill plus increased Raman scattering on those beams. These problems are mitigated in low gas fill hohlraums and this tactic has been successfully applied to plastic\cite{Kritcher2018} and HDC\cite{Zylstra2021} ablator designs.}. 

Second, the `I-Raum' campaign explored a change to the shape of the hohlraum,\cite{Robey2018} shown in Figure \ref{setup}, which has a larger hohlraum radius at the north and south poles of the target.  The I-Raum is cylindrically symmetric with a larger radius where the outer beam cones hit the wall, which radially displaces the high-$Z$ bubble generated by the outer beams ablating the hohlraum wall\footnote[2]{This high-Z plasma bubble has been shown to absorb inner beam energy before it reaches the equator of the target, significantly perturbing the symmetry of the x-ray drive on the capsule\cite{doi:10.1063/1.5023008}. For a given pulse length, this displacement improves the mode-2 drive symmetry. Modifying the shape of the hohlraum allows a more uniform drive without the loss of radiation temperature that simply moving to a larger hohlraum would entail, since the peak radiation temperature $T_{r}\sim(E_{Laser}/A_{Hohl})^{0.3}$ (Ref. \citenum{doi:10.1063/1.871025}).}. The I-Raum campaign also uses $\Delta \lambda$ for additional symmetry control. Each tactic for controlling $P_2$ symmetry was initially tested on a series of non-DT surrogate experiments that measure the symmetry at different phases of the implosion to validate the technique. 

Four experiments are reported here: N201101 and N210207 in the HyE campaign, and N201122 and N210220 in the I-Raum campaign [NIF shot numbers are denoted NYYMMDD where YY is the year, MM is the month, and DD is the date of the shot]. Differences between these four shots are discussed later. The targets used here are summarized in Fig. \ref{fig:setup}a-b, the design is discussed in detail in Ref. \citenum{KritcherYoungRobeyBurningPlasmaNatPhys}. Days before the shot the target is installed in the cryogenic positioner, filled with fuel and ice layered, aligned relative to the laser, and then shot with the laser pulse shapes designed to create a particular $T_r$ time history, peaking $\sim 300$ eV(see Fig. \ref{fig:setup}c). See the Methods for a full description of the preceding campaigns and details on the target fabrication, quality control, and experimental operations. 

\begin{table*}[t]
	\caption{Key data and inferences for the four experiments. Some observables are measured by neutrons (n), x rays (x), and/or $\gamma$ rays ($\gamma$).}
	\begin{center}
		\def\arraystretch{1.1}
		\begin{tabular}{|c|c|c|c|c|}
			 & N201101 (HyE) & N201122 (I-Raum) & N210207 (HyE) & N210220 (I-Raum) \\ \hline
			Total yield (Y$_{total}$, $\times 10^{16}$) & $3.49 \pm 0.10$ & $3.77 \pm 0.12$ & $6.07 \pm 0.17$ & $5.70 \pm 0.15$ \\
			~~~(kJ) & $98.4 \pm 2.7$ & $106.1 \pm 3.4$ & $171.0 \pm 4.8$ & $160.6 \pm 4.2$ \\ \hline 
			DT ion temperature ($T_{i,DT}$, keV) & $4.95 \pm 0.12$ & $5.17 \pm 0.13$ & $5.66 \pm 0.13$ & $5.54 \pm 0.14$ \\ \hline
			DD ion temperature ($T_{i,DD}$, keV) & $4.61 \pm 0.14$ & $4.65 \pm 0.14$ & $5.23 \pm 0.16$ & $5.13 \pm 0.24$ \\ \hline
			$4\pi$ down scattered ratio (DSR, $\%)$ & $3.44 \pm 0.16$ & $3.33 \pm 0.14$ & $3.16 \pm 0.16$ & $3.31 \pm 0.14$ \\ \hline
			Neutron $P_{0,n}$ ($\mu$m) & $38.5 \pm 1.1$ & $38.9 \pm 1.1$ & $42.3 \pm1.1$ & $37.6 \pm3.0$ \\ \hline
			Neutron $P_{2,n}$ ($\mu$m) & $8.0 \pm 0.3$ & $-5.1 \pm 0.8$ & $2.7 \pm 0.2$ & $-2.8 \pm 0.4$ \\ \hline
			Neutron $M_{0,n}$ ($\mu$m) & $33.6 \pm 1.0$ & $40.7 \pm 1.0$ & $39.3 \pm 1.0$ & $40.6 \pm 1.0$ \\ \hline
			X-ray $P_{0,x}$ ($\mu$m) & $39.2 \pm 2.6$ & $39.0 \pm 1.7$ & $48.0 \pm 2.3$ & $41.9 \pm 1.4$ \\ \hline
			X-ray $P_{2,x}$ ($\mu$m) & $15.7 \pm 0.7$ & $3.9 \pm 1.2$ & $6.3 \pm 0.9$ & $-1.4 \pm 0.8$ \\ \hline
			X-ray $M_{0,x}$ ($\mu$m) & $41.2 \pm 1.4$ & $54.3 \pm 3.4$ & $53.2 \pm 1.4$ & \textit{n/a} \\ \hline
			Gamma bang time (BT$_\gamma$, ns) & $9.37 \pm 0.03$ & $8.67 \pm 0.03$ & $9.09 \pm 0.02$ & $8.79 \pm 0.03$ \\ \hline
			Gamma burn width (BW$_\gamma$, ps) & $141 \pm 30$ & $150 \pm 20$ & $137 \pm 30$ & $139 \pm 20$ \\ \hline
			X-ray bang time (BT$_x$, ns) & $9.37 \pm 0.05$ & $8.69 \pm 0.05$ & $9.13 \pm 0.04$ & $8.79 \pm 0.04$ \\ \hline
			X-ray burn width (BW$_x$, ps) & $116\pm 6$ & $133 \pm 7$ & $99 \pm 6$ & $134 \pm 7$ \\ \hline
			Mode 1 drift velocity ($v_{p1}$, km/s) & $55 \pm 11$ & $100 \pm 12$ & $73 \pm 12$ & $123 \pm 10$ \\ \hline
			Inferred implosion velocity ($v_{imp}$, km/s) & $385 \pm 10$ & $380 \pm 10$ & $389 \pm 10$ & $374 \pm 10$ \\ \hline
			Coast time ($t_{coast}$, ns) & $0.92$ & $0.99$ & $0.87$ & $1.10$\\ \hline
			Peak $T_r$ (eV) & $301 \pm 8$ & $304 \pm 8$ & $307 \pm 8$ & \textit{n/a} \\ \hline
			$T_r$ 500ps before BT (eV) & $270 \pm 7$ & $256 \pm 6$ & $288 \pm 7$ & \textit{n/a} \\ \hline
			Mix Fraction ($MF$) & $0.44 \pm 0.06$ & $0.37 \pm 0.04$ & $0.47 \pm 0.06$ & $0.44 \pm 0.05$ \\ \hline
		\end{tabular}
	\end{center}
	\label{tab:data}
\end{table*}%

% Key data from these four experiments are given in Table \ref{tab:data} including the total fusion neutron yield (Y$_{total}$ as a total number and in kJ), ion temperature ($T_i$), down-scattered ratio (DSR)\cite{Frenje2013}, mean radii ($P_0$ and $M_0$ from equatorial and polar views) and Legendre mode 2 amplitude ($P_2$), bang time (BT), burn width (BW), mode 1 drift velocity ($v_{p1}$), inferred implosion velocity ($v_{imp}$), coast time ($t_{coast}$), and emission mix fraction ($MF$). 
Data from these four experiments are given in Table \ref{tab:data} (see left column for definitions of terms). A full suite of diagnostics\cite{doi:10.13182/FST15-173} record data on each experiment to characterize the shot's performance. A robust suite of diagnostics, developed over the past decade or more for NIF, is crucial for understanding these experiments and enabling adequate interpretation even as some systems, such as time-resolved x-ray imaging, became inoperable at higher yields. The key data used in inferring hot-spot conditions which are used in the burning plasma analysis (see Methods of Ref. \citenum{ZylstraBurning}) are Y$_{total}$, $T_i$, volume, which is inferred from the mean radii and Legendre amplitudes, and BW. For the Hurricane burning-plasma criteria\cite{doi:10.1063/1.5087256}, $v_{imp}$ is also required. Additional quantities in Table \ref{tab:data} are key inputs for constraining the simulation models of these experiments (e.g. the BT and DSR), or are used to understand the empirical scaling of these results ($t_{coast}$). Each critical input for the burning-plasma analysis is measured with redundancy of detectors and/or measurement methodology. 

The absolute unscattered neutron yield is measured using neutron activation detectors (NAD)\cite{doi:10.1063/1.4733741} and the magnetic recoil spectrometer (MRS);\cite{Johnson2012} the final value is a weighted mean. The NAD measures neutron activation of three redundant zirconium samples to infer the yield based on the activity, known activation cross section, and sample mass and solid angle. The MRS measurement is independent and uses elastic scattering of deuterons in a foil close to the target, momentum analyzed using a dipole magnet; the systematic uncertainty is primarily the deuteron scattering cross section and an empirical calibration of the charged-particle collection efficiency. Each measurement also includes a statistical uncertainty, on these shots the two independent measurements agree with a typical difference of $\sim 7\%$, consistent with the $\sim 5\%$ uncertainty on each measurement. Y$_{total}$ includes a correction for the fact that some fraction of the neutrons scatter in the dense shell\cite{doi:10.1063/5.0043589}, the ratio is $e^{4 \times DSR}$. This few percent uncertainty in yield propagates directly into all the burning-plasma criteria.\cite{ZylstraBurning}

$T_i$ is inferred from the 2$^{\textrm{nd}}$ moment of the neutron energy spectrum\cite{ballabio1998relativistic}, measured using neutron time of flight (NTOF) detectors\cite{Glebov2010,Hatarik2015} which report an $T_i$ from both the DT and DD fusion reactions, the latter is less affected by residual kinetic energy (RKE)\cite{Murphy2014} in the burning fuel; the MRS also reports a DT $T_i$. Between 3-5 NTOFs at different lines of sight recorded good data for $T_{i,DD}$ and $T_{i,DT}$ on each shot and are self-consistent, the primary uncertainty results from the temporal instrument response function which is measured using ultra-short impulse shots. Electron temperature measurements are not available on all shots but for N210207, the highest performer, a differentially-filtered measurement\cite{PhysRevE.101.033205} gives $T_e = 5.19 \pm 0.14$ which is consistent with $T_{i,DD}$ as expected.\cite{PhysRevLett.121.085001}

The mean hot-spot radius ($P_0$, $M_0$) and $P_2$ are measured with neutron\cite{Volegov2014} and x-ray\cite{doi:10.1063/1.4733313} imaging; at the equator, these measure polar modes ($P_0$ and $P_2$) and from the pole, they measure the mean radius in the equatorial plane ($M_0$). In both cases the measurement technique relies on either pinhole or penumbral imaging with the aperture placed close to the implosion. The x-ray detector is image plate with filtering that results in a measurement $\sim 15$ keV photon energy. The neutron imaging uses a redundant system of image plates and time-gated cameras to record images at different neutron energy ranges. Here we quote only $P_0$, $M_0$, and $P_2$, which are the dominant low-mode asymmetries measured in imaging diagnostics and the most important for understanding the shot performance and inferring a total volume. The measured x-ray volumes are systematically larger than neutron volumes, this offset is expected and depends on the photon energy predominantly measured in the imaging system. Since the hot-spot analysis methodologies\cite{doi:10.1063/5.0003298} are validated using neutron-measured volumes these are used in the burning-plasma analysis in Ref. \citenum{ZylstraBurning}.

BT is defined as the time of peak emission while BW is defined as the full-width at half-max (FWHM) of the emission. Nuclear BT and BW are measured by the $\gamma$ reaction history instrument (GRH)\cite{Herrmann2010}, in which incident $\gamma$ rays Compton scatter electrons into a pressurized gas cell where they exceed the medium's speed of light, generate Cherenkov light, which is recorded by a photomultiplier tube. X-ray bang time and burn width are measured using SPIDER\cite{Khan2012}, which uses a set of differing attenuation filters with a slit and streak camera to record the x-ray emission history at varying photon energies. The temporal response of each instrument, and its absolute timing, is measured using laser-generated impulses and deconvolved from the measured signals to infer the BT and BW. These two measurements are very consistent with each other, especially in BT. The measured BW$_x$ may be systematically smaller due either to an instrumental effect or physics, as they measure differently-weighted burn histories and the x-ray history is known to be dependent on the photon energy; the largest discrepancy is the BW on N210207 in which the difference is a 1.2$\sigma$ event with normally-distributed uncertainties, which is not unlikely ($\sim 11\%$ likelihood). 

\begin{figure*}[t]
	\centering
	\includegraphics[width=4.75in]{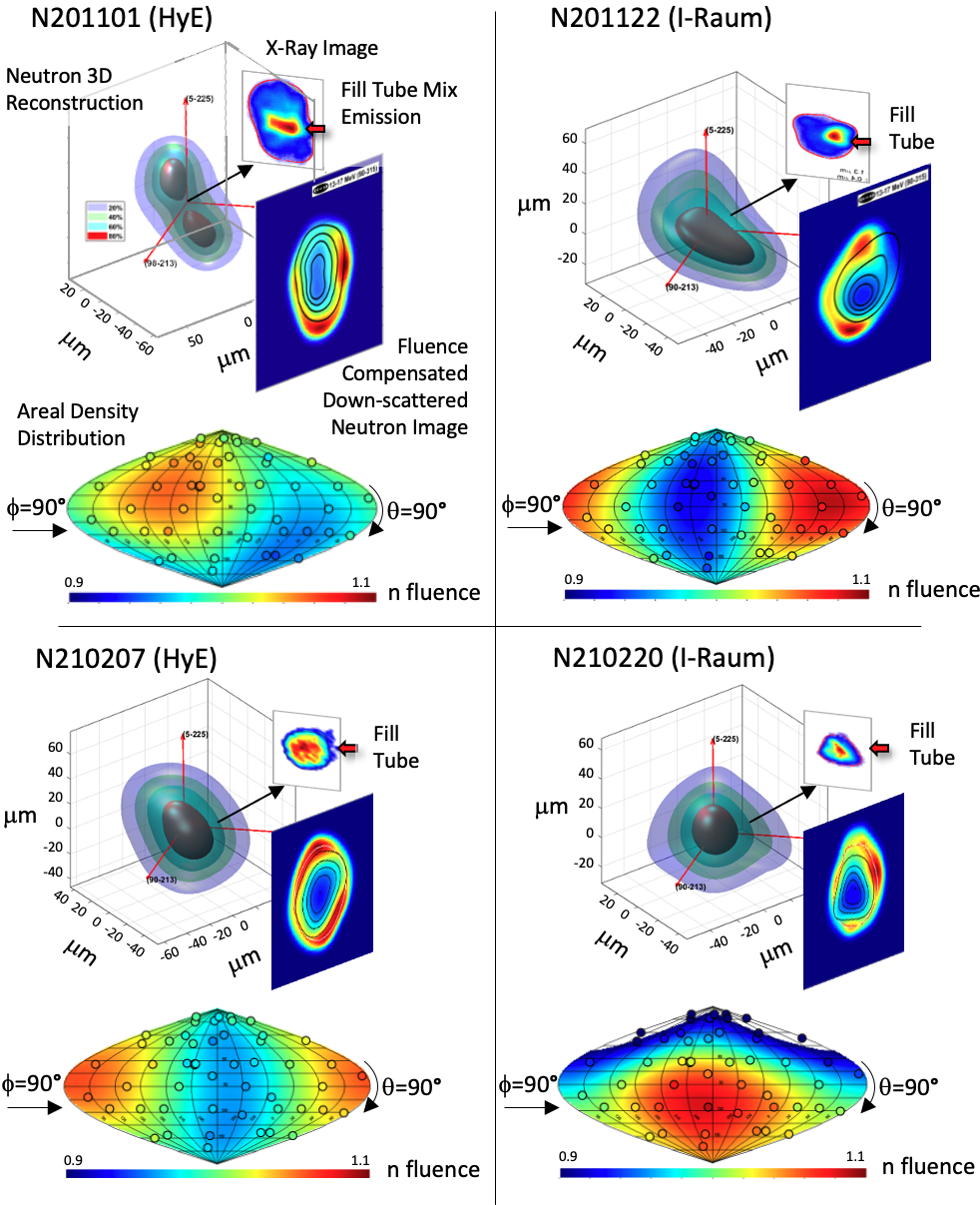}
	\caption{Detailed measurements on the implosion symmetry for the shots, from top left: N201101, N201122, N210207, and N210220. At top shows a reconstruction of the neutron emissivity in 3-D. X-ray self-emission images are taken along the equator (upper right). At the lower right of the neutron emissivity reconstruction is a fluence-compensated scattered neutron image, representing the dense shell distribution. At bottom is an Aitoff projection of the measured neutron fluence, inversely related to the shell $\rho R$, with a color scale from $0.9-1.1$; in these images the azimuthal angle $\phi=90^\circ$ is at the left side of the image increasing to the right, and the polar angle $\theta$ increases from top to bottom with $90^\circ$ at the center.}
	\label{fig:symmetry}
\end{figure*}

DSR is calculated from the neutron spectra measured by NTOFs and MRS, and is defined as the ratio of the integrated neutron spectrum between 10-12 MeV and 13-15 MeV. The former represent neutrons which undergo a single elastic scattering in the compressed fuel, so the ratio is then related to the total areal density ($\rho R$) of the fuel.\cite{Frenje2013} Since each detector samples only a small fraction of the shell, and thus can be biased by 3-D asymmetries, a `$4 \pi$' DSR, which is given in Table \ref{tab:data}, is calculated by fitting a model to all of the available data (5 or 6 lines of sight depending on the specific experiment).\cite{doi:10.1063/5.0043589}

The mode 1 drift velocity ($v_{p1}$) is measured using time-of-flight detectors\cite{Moore2018} which precisely measure the difference in arrival time between photons and the main neutron peak across five lines of sight to infer the bulk velocity of the neutron-emitting plasma. $v_{imp}$ is inferred from the bang time using a rocket model and relevant trajectory measurements on non-DT shots using the methodology described in Refs. \citenum{doi:10.1063/1.3592170,doi:10.1063/1.4803915}. The time-dependent $T_r$ is measured using the DANTE\cite{Kline2010} instrument, which is an eighteen-channel x-ray spectrometer with a variety of calibrated transmission filters and x-ray mirrors coupled to x-ray diodes; $t_{coast}$ is taken as the difference in time between $T_r$ falling to 95\% of its peak value and BT. Table \ref{tab:data} gives peak $T_r$ values and $T_r$ 500ps before BT; DANTE did not acquire data for N210220. Lastly, $MF$ is a representation of the amount of x-ray emission due to high-$Z$ mix, or capsule material, that is injected into the hot spot during burn. High-$Z$ material enhances bremsstrahlung losses from the plasma, and $MF$ is extracted from the high-frequency components of the x-ray image data using the same prescription as in Ref. \citenum{Pak2020}. Note that this MF is for high-energy ($> 10$ keV) radiation, an approximate relation to convert this into a bremsstrahlung power ($P_b$), where $> 2$ keV photons are relevant, is $P_b \sim (1+0.4 MF)P_{b,DT}$ (see Supplemental Material of Ref. \citenum{Pak2020}), where $P_{b,DT}$ would be the bremsstrahlung power for a pure DT plasma, so that a $MF \sim 0.4-0.5$ corresponds to a radiation loss enhancement of $0.16-0.20$. Previous work indicates that this mix arises from hydrodynamic growth of perturbations seeded by the capsule fill tube and capsule manufacturing defects (see Methods and Ref. \citenum{Zylstra2020}).

\begin{figure}[t!]
\centering
\includegraphics[scale=1]{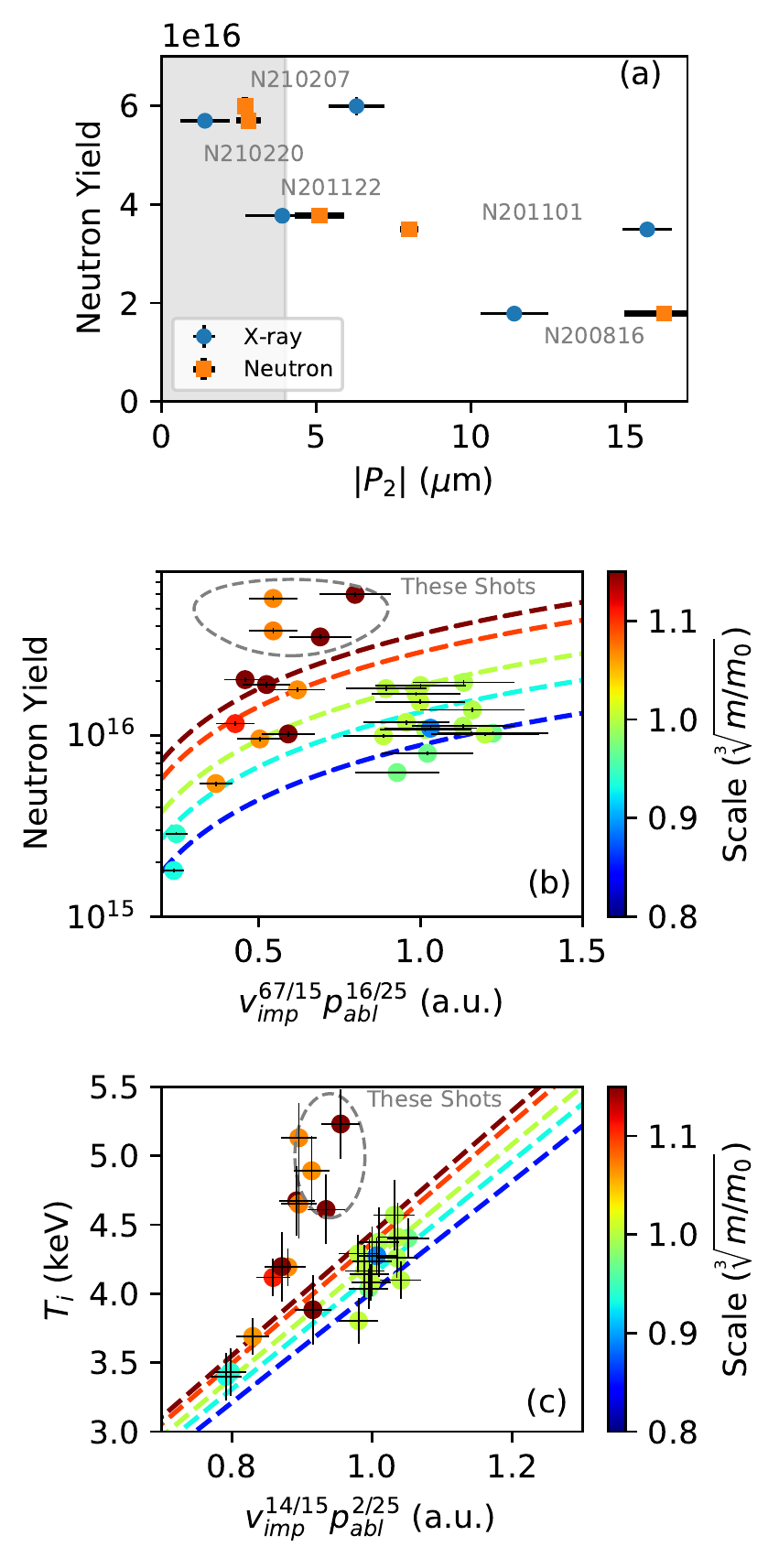}
\caption{\label{fig:scaling} Performance scaling of these implosions. (a) Yield versus $|P_2|$, measured from both x-ray and neutron imaging (see Table \ref{tab:data}, for shots with similar implosion parameters, demonstrating a strong scaling with the high-performing implosions having the lowest $|P_2|$. (b) Yield versus a scaling parameter based on $v_{imp}$ and the in-flight ablation pressure $p_{abl}$, with previous data from NIF and this work color coded by scale of the implosion. The dashed curves represent the expected scaling with a modest level of self-heating and constant in-flight adiabat, degradations, and residual kinetic energy. (c) Similar plot as (b) except with $T_{i,DD}$ plotted against the expected scaling for modest levels of self-heating.}
\end{figure}

Detailed images of these four implosions are shown in Fig.~\ref{fig:symmetry}, including a 3-D reconstruction of the neutron emissivity\cite{Volegov2014}, fluence-compensated scattered neutron image, which represents the dense DT shell distribution\cite{Casey2016}, x-ray self-emission image, and Aitoff projection of measurements of the relative neutron fluence, which is inversely related to the shell $\rho R$.\cite{Bionta_RSI_2021} Some notable low-mode asymmetries remain in these implosions: N201101 has a large positive $P_2$ (prolate) asymmetry, resulting in two distinct lobes of high neutron emissivity while the fluence-compensated image shows a low $\rho R$ at the north pole, due to the coherent addition of a $\sim 55$ km/s mode 1 velocity (see Table \ref{tab:data}) along that direction\cite{Rinderknecht2020}. The fluence distribution also shows an asymmetry in the equatorial plane. Shot N201122 has a low $P_2$ seen in both the neutron reconstruction and x-ray images, but a larger mode 1 asymmetry, notable in both the reconstructed shell density image, the neutron fluence distribution, and in a $\sim 100$ km/s hot-spot velocity (Table~\ref{tab:data}). This mode-1 asymmetry was caused by a known asymmetry in the shell thickness on this experiment, at the time the sensitivity of our low-mode asymmetry to these initial seeds was being developed\cite{Casey2021}. The third experiment, N210207, was based on N201101 and reduced the amount of $\Delta \lambda$ to reduce the $P_2$ asymmetry which is evident in the data, with the low-mode symmetry improved. The fourth experiment, N210220, used an imposed laser flux asymmetry to compensate for the known mode 1 asymmetry sources, which will be detailed in an upcoming publication; the diagnostics still show residual mode 1 albeit with less of an obvious hole in the shell areal density. All experiments show a bright feature in the x-ray emission images corresponding to the fill tube\cite{Pak2020}. 

There were two key aspects of implosion control that enabled the successful creation of these burning plasmas. First, controlling the low-mode symmetry is crucial to the high performance.\cite{KritcherYoungRobeyBurningPlasmaNatPhys} Without either tactic for symmetry control - using $\Delta \lambda$ to adjust transfer or introducing the pocket - these implosions would have been highly oblate, with $|P_2|$ up to $\sim 50$ $\mu$m\cite{doi:10.1063/1.5020057,Zylstra2021}. Fig. \ref{fig:scaling}a shows the measured fusion yield versus the mode-2 asymmetry magnitude ($| P_2|$) for a set of shots in these two campaigns with reasonably comparable implosion parameters, and a strong sensitivity to $P_2$ is observed. For example in the I-Raum campaign, from the earlier shot N200816 to N201122 the only change was an application of 0.5\AA$ $ of $\Delta \lambda$ (see Methods) which substantially reduced the asymmetry and roughly doubled the yield. Similarly in HyE, one of the main changes from N201101 to N210207 was an improvement in $P_2$ which contributed to the improved performance. The gray shaded region in Fig. \ref{fig:scaling} corresponds to the 4 $\mu$m specification from Ref. \citenum{Kritcher2014}, we see that maintaining the symmetry close to this specification appears key for high performance.

Implosion performance also displays highly sensitive dependence on other implosion parameters. In a regime of modest self-heating, Hurricane et al. published\cite{ISI:000451447200001} the following relationship for the fusion yield and temperature:
\begin{eqnarray} 
	Y_{DT} &\propto& p_{abl}^{16/25} \frac{v_{imp}^{67/15}}{\alpha_{if}^{36/25}}S^{14/3} \eta \left(1-RKE \right)^{23/7}
	\label{eqn:Yscaling}
	\\
	T_i &\propto& \frac{p_{abl}^{2/25} v_{imp}^{14/15}}{\alpha_{if}^{9/50}} S^{1/3}
	\label{eqn:Tscaling}
\end{eqnarray}
where $v_{imp}$ is the implosion velocity defined earlier. The other parameters are the in-flight ablation pressure ($p_{abl}$) which represents the drive pressure acting on the imploding shell $\sim 500$ ps before bang time, a higher $p_{abl}$ prevents in-flight decompression of the fuel. We empirically assess this from DANTE measurements of $T_r(t)$ with the ablation-pressure relationship\cite{Landen2021} $p_{abl} \propto T_r^{2.4}$. $\alpha_{if}$ is the in-flight `adiabat', which represents the entropy, or compressibility, of the DT fuel and is defined as the ratio of the fuel's pressure at peak velocity to its Fermi pressure. $S$ is the scale factor of the implosion, simplistically this can be taken as the initial radius, here we use $\sqrt[3]{M_{DT}}$ where $M_{DT}$ is the fuel mass. The remaining terms represent degradation mechanisms: $\eta = (1-MF)$ is a measure of the fuel's cleanliness,\cite{Pak2020} decreasing with mix or enhanced bremsstrahlung emission, and $RKE$ is the residual kinetic energy at stagnation compared to the peak implosion kinetic energy. The latter is a measure of the wasted implosion kinetic energy that is not converted to heating of the fuel, and increases with low-mode asymmetry\cite{HurricanePiston2020POP}. 

Maintaining $\eta$ and $RKE$ as close to previous experiments, by controlling the sources of these degradations, was a key component of this work's success (see Methods). By selecting a set of previous HDC-ablator campaigns with similar $\alpha_{if}$ and modest degradations ($\eta$ and $RKE$) we can examine the performance scaling with the remaining parameters: $p_{abl}$, $v_{imp}$, and $S$. Fig. \ref{fig:scaling}b shows the yield data plotted against $p_{abl}^{16/25} v_{imp}^{67/15}$ (see Eq. \ref{eqn:Yscaling}) and the shots color-coded by $S$. Contours of $S^{14/3}$ are shown by the dashed curves, and the abscissa is normalized by shot N170827 from Ref. \citenum{PhysRevLett.120.245003}. Similarly Fig. \ref{fig:scaling}c shows $T_{i,DD}$ plotted versus normalized $p_{abl}^{2/25} v_{imp}^{14/15}$ with contours of $S^{1/3}$. In the modest-self-heating regime which this relationship was derived for, we would expect the data to follow the curve corresponding to their color ($S$), some scatter is observed due to variability of the degradation mechanisms. These higher-performing implosions, in shades of orange to red, agree with the expected scaling at low performance (low $p_{abl}^{16/25} v_{imp}^{67/15}$ or $p_{abl}^{2/25} v_{imp}^{14/15}$). In the experiments reported here, we successfully increased this parameter and we see that the large-scale implosions follow a rapid increase in both yield and temperature, faster than expected from Eqs. \ref{eqn:Yscaling} and \ref{eqn:Tscaling} because of the high levels of self-heating; new scaling relationships need to be derived for the burning-plasma regime. From this we conclude both that the strategy to increase $S$ was successful [as these implosions have significantly higher yield (temperature) than the smaller-scale implosions, even at slightly less $p_{abl}^{16/25} v_{imp}^{67/15}$ ($p_{abl}^{2/25}v_{imp}^{14/15}$)], and that further increases in $p_{abl}$ and $v_{imp}$, e.g. by continued improvements in hohlraum efficiency, are expected to continue increasing performance. 

In summary, two related experimental platforms have improved the performance of DT layered implosions on the NIF by a factor of $3 \times$ in yield over previous results, into a new burning plasma regime\cite{ZylstraBurning} where alpha heating is now the dominant contributor to energy in the hot spot. The data measured on these four shots, which support the burning-plasma conclusion of Ref. \citenum{ZylstraBurning} and constrain the computation models discussed in Ref. \citenum{KritcherYoungRobeyBurningPlasmaNatPhys}, are presented. The increased performance represents the culmination of a strategy to increase the implosion scale, which is expected to rapidly increase performance (Eq. \ref{eqn:Yscaling}) but is a necessary, but not sufficient, condition for actually realizing the improvement. These experiments resulted in record performance with peak fusion power exceeding peak incident laser power by a factor of 2x because we were able to implement tactics for controlling the low-mode symmetry of the implosion in an aggressive regime\cite{doi:10.1063/1.5020057}, increase other key implosion performance parameters from previously-reported large-scale implosions\cite{Hohenberger2020,Zylstra2021}, and control degradation mechanisms. These conclusions are supported by published scaling relationships for implosion performance; these experiments demonstrate higher performance than previous data due to the higher sensitivity in the burning-plasma regime. While degradations were controlled to the level where they did not preclude this level of performance, further reduction in enhanced radiation losses from high-$Z$ mix and improvements in low-mode symmetry are possible and expected to lead to performance improvements. Similarly, further increases in drive coupling to the capsule, for example from more efficient hohlraums, is also projected to lead to continued performance improvements. These will be a focus of the experimental program on NIF going forward, as well as conducting experiments to study physics in the burning-plasma regime, such as using `dudded', low-alpha-heating fuel (THD instead of DT) to directly explore the effects that fusion self-heating is having on the plasma.

%\clearpage
{\bf Acknowledgments and Disclaimer} We thank John Kline (LANL) and Mike Farrell (GA) for thoughtful discussions. The contributions of NIF operations and target fabrication teams to the success of these experiments are gratefully acknowledged. 
This work performed under the auspices of the U.S. Department of Energy under Contract No. DE-AC52-07NA27344 and Contract no. 89233218CNA000001.       
This document was prepared as an account of work sponsored by an agency of the United States government. Neither the United States government nor Lawrence Livermore National Security, LLC, nor any of their employees makes any warranty, expressed or implied, or assumes any legal liability or responsibility for the accuracy, completeness, or usefulness of any information, apparatus, product, or process disclosed, or represents that its use would not infringe privately owned rights. Reference herein to any specific commercial product, process, or service by trade name, trademark, manufacturer, or otherwise does not necessarily constitute or imply its endorsement, recommendation, or favoring by the United States government or Lawrence Livermore National Security, LLC. The views and opinions of authors expressed herein do not necessarily state or reflect those of the United States government or Lawrence Livermore National Security, LLC, and shall not be used for advertising or product endorsement purposes.  LLNL-JRNL-821520-DRAFT.
\\

% ------------------------------------------------ %
%  Author Contributions
% ------------------------------------------------ %

\section*{Author contributions}
\leavevmode \\
J.S.R.~I-Raum experimental lead, N201122\&N210220 ``Shot RI'' (shot responsible individual), and wrote sections of the paper; J.E.R.~N201101 \& N210207 experimentalist and Shot RI, wrote sections of the paper; A.B.Z.~hot-spot analysis lead, Hybrid-E experimental lead, wrote sections of the paper; A.L.K.~Hybrid-E design lead, integrated hohlraum group lead; H.F.R.~original I-Raum design lead; C.V.Y.~present I-Raum design lead; O.A.H.~capsule scale/burning plasma strategy, theory, 0D hot-spot models; D.A.C.~empirical hohlraum $P_2$ model and hohlraum strategy; K.L.B.~Hybrid Shot RI; D.T.C.~Hybrid Shot RI; T.D.~Hybrid Shot RI; L.D.~3D hot-spot analysis; M.H.~Hybrid Shot RI; S.L.P.~Hybrid Shot RI; A.P.~Hybrid \& I-Raum Shot RI, physics of capsule engineering defects; P.K.P.~1D hot-spot analysis, $Y_{\rm{amp}}$ and GLC inference; R.T.~Hybrid Shot RI; S.J.A.~capsule microstructure physics; P.A.A.~hohlraum physics; L.J.A.~engineering and targets; B.B.~penumbral x-ray diagnostic; D.B.~computational physics; L.R.B.~x-ray framing camera; L.B.H.~HDC design and campaign lead; R.B.~ICF physics/ignition theory; S.D.B.~cryo layering; R.M.B.~RTNAD nuclear diagnostic; N.W.B.~neutron diagnostics; E.J.B.~project engineering; D.K.B.~diagnostics; T.B.~capsule fab \& metrology; T.M.B.~cryo layering; M.W.B.~project engineering; P.M.C.~DT EOS measurements; B.C.~HYDRA code development; T.C.~LPI physics; H.C.~GLEH x-ray diagnostic; C.C.~target fab planning; A.R.C.~ignition theory; D.S.C.~capsule/instability physics; J.W.C.~capsule fabrication; E.L.D.~experiments; T.R.D.~capsule physics; M.J.E.~program management; W.A.F.~hohlraum physics; J.E.F.~2DConA image analysis; D.F.~nuclear diagnostics; J.F.~magnetic recoil spectrometer nuclear diagnostic; J.G.~ensemble simulations; M.G.J.~magnetic recoil spectrometer diagnostic; S.H.G.~ICF physics; G.P.G.~nuclear diagnostics; S.H.~capsule physics, iPOM analysis; K.D.H.~neutron diagnostics; G.N.H.~experiments; B.A.H.~capsule physics; J.H.~computational physics; E.H.~nuclear time-of-flight diagnostics; J.E.H.~MOR and PAM stability, SSD improvements, and FC control; V.J.H.~MOR and PAM stability, SSD improvements, and FC control; H.H.~gamma diagnostics; M.C.H.~program management; D.E.H.~hohlraum physics, CBET studies in Hybrid-C; D.D.H.~capsule physics; J.P.H.~x-ray diagnostics; W.W.H.~management; H.H.~capsule fabrication; K.D.H.~ensemble simulations; N.I.~x-ray diagnostics; L.J.~x-ray diagnostics; J.J.~neutron diagnostics; O.J.~hohlraum physics; G.D.K.~HYDRA code development; S.M.K.~neutron diagnostics; S.F.K.~x-ray diagnostics and analysis; J.K.~diagnostic management; Y.K.~gamma diagnostics; H.G.K.~gamma diagnostics; V.G.K.~neutron diagnostics; C.K.~capsules; J.M.K.~HYDRA code development; J.J.K.~targets; O.L.L.~velocity analysis; S.L.~laser plasma instability (PF3D) code development; D.L.~NIF facility management; N.C.L.~optical diagnostics; J.D.L.~ICF physics; T.M.~ICF physics; M.J.M.~x-ray diagnostics; B.J.M.~mode-1 analysis, backscatter; A.J.M.~diagnostic management; S.A.M.~integrated design physics; A.G.M.~x-ray diagnostics; M.M.M.~HYDRA code development lead; D.A.M.~x-ray diagnostics; E.V.M.~x-ray diagnostics; L.M.~capsule physics; K.M.~gamma diagnostics; N.B.M.~hohlraum physics; P.A.M.~LPI physics; M.M.~optical diagnostics; J.L.M.~hohlraum physics; J.D.M.~hohlraum physics; A.S.M.~neutron diagnostics; J.W.M.~hohlraum physics; T.M.~neutron and gamma diagnostics; K.N.~project engineering; J.G.D.N.~MOR and PAM stability, SSD improvements, and FC control; A.N.~target fab engineering, capsule, and fab planning; R.N.~ensembles simulations; M.V.P.~HYDRA code development; L.J.P.~MOR and PAM stability, SSD improvements, and FC control; J.L.P.~ensembles simulations; Y.P.~hohlraum physics; B.B.P.~hohlraum physics; M.R.~capsule fabrication; N.G.R.~capsule fabrication; H.R.~RTNAD mode-1 analysis; M.R.~hohlraum physics; M.S.R.~x-ray diagnostics; J.D.S.~hohlraum physics; J.S.~mode-1 analysis; S.S.~capsules; D.J.S.~neutron diagnostics; M.B.S.~hohlraum diagnostics; C.R.S.~HYDRA code development; H.A.S.~NLTE opacities (Cretin) code development; S.M.S.~HYDRA code development; K.S.~mode-1 metrology; M.W.S.~kinetic physics; S.S.~sagometer data \& particle analysis; V.A.S.~capsule physics; B.K.S.~ensemble simulations; P.T.S.~dynamic model, ignition theory; M.S.~capsules; S.S.~x-ray diagnostics; D.J.S.~hohlraum/LPI physics; L.J.S.~hohlraum physics; C.A.T.~Bigfoot design physics; R.P.J.T.~program management; E.R.T.~optical diagnostics; P.L.V.~neutron imaging diagnostics;  C. R. W. capsule/instability physics; K.W.~x-ray diagnostics; C.W.~capsule fabrication; C.H.W.~neutron diagnostics; B.M.V.W.~NIF operations lead; D.T.W.~hohlraum physics; B.N.W.~project engineering; M.Y.~capsule fabrication; S.T.Y.~MOR and PAM stability, SSD improvements, and FC control; G.B.Z.~computational physics lead.

\textbf{Competing Interests} The authors declare that they have no competing financial interests.

\textbf{Correspondence} Correspondence and requests for materials
should be addressed to J.S.R., J.E.R., and A.B.Z. (emails: ross36@llnl.gov, ralph5@llnl.gov, zylstra1@llnl.gov).

\bibliographystyle{naturemag}
\bibliography{BibFile2}

\clearpage
\section*{Methods}
\subsection{Campaign Summary}
The Hybrid E campaign originally began with larger radius (1100 $\mu$m) capsules, which achieved the previous record for implosion energetics and fusion yield ($\sim 56$ kJ) \cite{Zylstra2021}. The shots published in Ref. \citenum{Zylstra2021} were limited to implosion velocities below $\sim 370$ km/s and had coast times $\sim 1.3$ ns. Longer coast times are deleterious because the imploding shell can begin to decompress in flight\cite{doi:10.1063/1.4900621,doi:10.1063/1.4994856}. Initial attempts to increase the velocity of these implosions was limited by the quality of capsules fabricated to date at this scale\cite{Kritcher2021_PoP}. Relative to Ref. \citenum{Zylstra2021}, in this work we decreased the diameter of the laser entrance hole (LEH) to increase radiation drive on the capsule, and decreased the capsule radius by $50$ $\mu$m. The first shot at these conditions, N201011, reproduced the velocity and coast time of N191110 from Ref. \citenum{Zylstra2021} and had about half the fusion yield ($\sim 29$ kJ), due to the smaller scale and from some mix from large capsule defects. The second experiment used an extended duration laser drive to simultaneously increase the implosion velocity to $\sim 385$ km/s and decrease the coast time to $\sim 0.9$ ns (increasing $p_{abl}$), this shot (N201101) produced a stagnation pressure roughly double that of its predecessor N201011 with a record yield, at the time, of $\sim 100$ kJ. With the extended pulse, the $P_2$ symmetry is expected to become more oblate, on N201101 this was compensated with $+0.5$ {\AA} of $\Delta \lambda$ relative to N201011; this was an overcompensation due to model uncertainty resulting in a highly prolate shape on N201101. On the third experiment, N210207, this was reduced by $0.2$ {\AA} resulting in a $P_2$ shape much closer to round; a new capsule batch was also used on N210207, the velocity and coast time improved slightly due to laser delivery, and the combination of these changes resulted in a substantial increase in fusion yield to 170 kJ.

The I-Raum campaign has used a single size of capsule, $1000$ $\mu$m inner radius. The first experiment, N190217, used a capsule with very significant levels of defects (similar to the poor-quality capsules reported in Ref. \citenum{Zylstra2020}) resulting in high hot spot mix and low yield ($\sim 19$ kJ), consistent with previous studies. The second DT experiment (N191105) used a better quality capsule and a higher initial shock pressure to increase the fuel adiabat and improve stability properties. N191105 observed a nearly tripled yield (50 kJ) but had a mild $P_2$ asymmetry and large mode 1 asymmetry, hypothesized to be from the loss of four inner beams. A repeat shot with full beam participation on N200816 reduced the mode 1 of N191105 but had a larger than expected negative $P_2$ asymmetry and resulted in similar performance (50 kJ). The $P_2$ shape was tuned closer to round with the application of 0.5 {\AA} of $\Delta \lambda$, leveraging the sensitivity curve demonstrated by Hybrid E, on the fourth shot (N201122), with the yield more than doubling from its predecessor to $\sim 106$ kJ. Shot N201122 experienced a large mode 1 asymmetry, so on the next shot N210220 the incident laser power was adjusted to partially compensate for the known sources of mode 1, this shot produced a large increase in performance to $\sim 160$ kJ. In this paper, we focus only on the last two experiments in each campaign, which have produced $\gtrapprox 100$ kJ yields.

\subsection{Target Quality}
Execution of a successful NIF experiment begins with fabrication\cite{doi:10.13182/FST15-163} of the targets to be used, the design details are summarized in Fig. \ref{fig:setup}a and discussed in detail in Ref. \citenum{KritcherYoungRobeyBurningPlasmaNatPhys}. The Hybrid E target has a diameter of 6.40 mm and a length of 11.24 mm. The I-Raum target has a diameter of 6.72 mm at the outer beam location and a diameter of 6.20 mm at the inner beam location near the center of the target, with a total target length is 11.4 mm. The total hohlraum wall area is similar in both cases. The hohlraums are made of depleted uranium with a 0.7 $\mu$m gold overcoat, which are fabricated by deposition onto a mandrel, which can be diamond-turned to the precise shape for each hohlraum shape\cite{Hein2013}. Each component is screened for coating defects that could lead to drive asymmetries or create particles that flake off of the wall onto the capsule. The laser beams enter the target through the laser entrance holes (LEHs), which are 3.64 mm in diameter for both platforms. The hohlraums include thin plastic windows covering the LEH to seal the interior volume and a gas fill line which is used to introduce He gas, in both designs the hohlraum interior is filled with 0.3mg/cc of helium. 

The capsule, located at the center of the target, is HDC and includes a tungsten-doped layer to reduce `preheat' from higher-than-thermal x-ray emission from the hohlraum. The nominal total thickness ($\sim 80$ $\mu$m, see Fig. \ref{fig:setup}b) and optical depth of the doped layer are specified by the capsule design to optimize the implosion and stability properties\cite{KritcherYoungRobeyBurningPlasmaNatPhys}. The HyE campaign uses a larger capsule than the I-Raum, 1050 vs 1000 $\mu$m inner radius. The capsules are fabricated using a chemical vapor deposition process\cite{Biener2006}. After fabrication, the actual thickness (mass) of each layer is used to refine the pre-shot computational models\cite{KritcherYoungRobeyBurningPlasmaNatPhys}. The capsules are also all metrologized in depth to detect several kinds of defects that may be present, which allows a quality-control step to ensure the highest-quality capsules are used in the experiment, additional quality control steps are completed after the completed target is assembled.

Several aspects of target quality can affect an experiment. For low-mode symmetry, the currently-dominant seed from fabrication is mode-1 asymmetry in the shell thickness,\cite{Casey2021} which can result from the capsule coating process. At present the thickness asymmetry is measured for each capsule before target builds by two x-ray transmission imaging\cite{doi:10.13182/FST07-3} methodologies: contact radiography (CR), which measures the asymmetry from one view and is thus a random projection, and `Xradia', which combines multiple views to reconstruct the full asymmetry magnitude. The uncertainty is similar in both cases and the two measurements are combined using a Bayesian model (BM) to create a more accurate estimation of the asymmetry amplitude. The orientation from these measurements to the final assembled target is unknown, so we additionally measure both the asymmetry's amplitude and its orientation when the target is installed in the cryogenic target positioner (Ctps) via the three orthogonal x-ray views\cite{Parham2016}. A comparison of the of the measured mode-1 shell asymmetry amplitudes is shown in Table \ref{tab:m1}. The different imaging systems are in good agreement on the mode-1 amplitude. The fact that the shell used on N210220 had a substantial asymmetry, which was known before the experiment, motivated using a compensating laser asymmetry\cite{MacGowanHEDP} to balance it.

\begin{figure}
	\centering
	\includegraphics[width=3in]{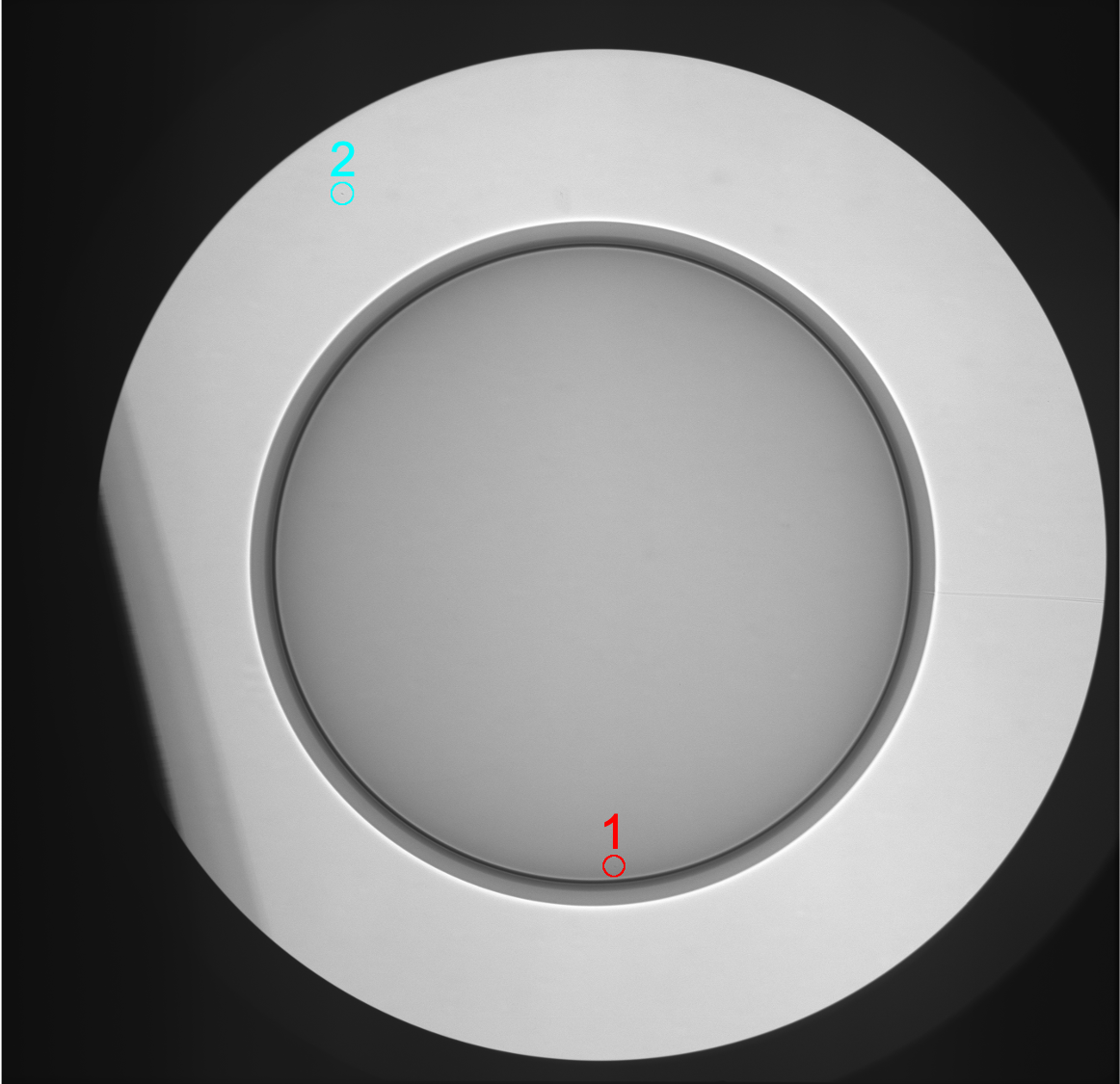}
	\caption{\label{fig:sagometer} Pre-shot x-ray radiograph of the capsule used on N210207 through the LEH, with the capsule at center. Two particles were detected via a parallax analysis.}
\end{figure}

\begin{figure*}[t]
	\centering
	\includegraphics[scale=1]{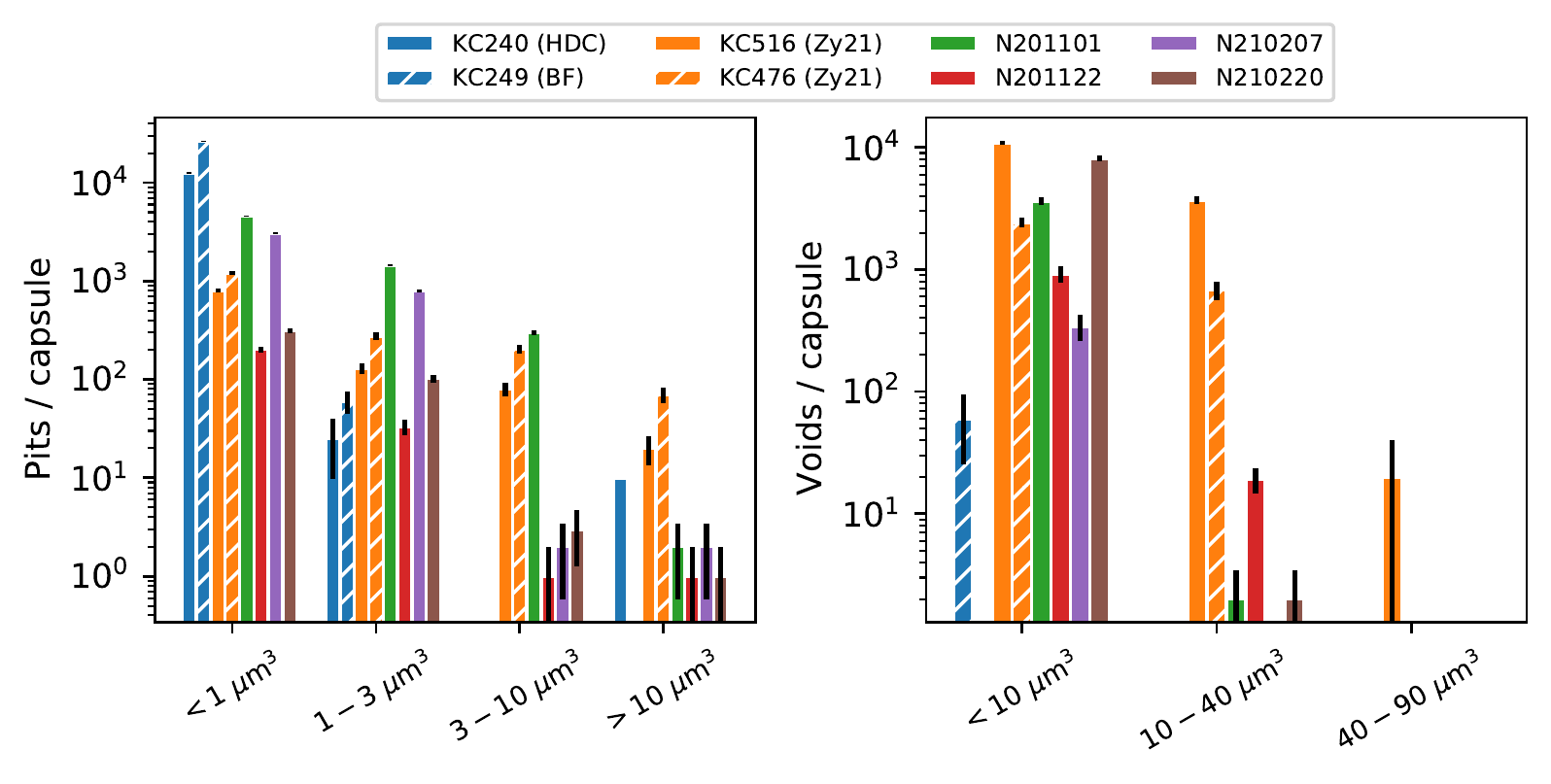}
	\caption{\label{fig:pitsvoids} Pits (left) and voids (right) per capsule with several bins by volume for capsule batches used in previous work with good quality (KC240, KC249; blue) and marginal quality (KC516, KC476; orange).}
\end{figure*}

\begin{table}[t]
	\caption{Measured capsule mode-1 asymmetry amplitudes via several complementary techniques}
	\begin{center}
		\small
		\def\arraystretch{1.2}
		\begin{tabular}{|c|c|c|c|c|}
			& N201101 & N201122 & N210207 & N210220 \\ \hline
			CR ($\mu$m, $\pm 0.2$) & 0.27 & 0.44 & 0.10 & 0.55
			\\ \hline
			Xradia ($\mu$m, $\pm 0.2$) & 0.12 & 0.34 & 0.16 & 0.47
			\\ \hline 
			BM ($\mu$m) & $0.22^{+0.14}_{-0.13}$ & $0.41^{+0.15}_{-0.15}$ & $0.18^{+0.14}_{-0.11}$ & $0.55^{+0.15}_{-0.15}$
			\\ \hline
			Ctps ($\mu$m, $\pm 0.21$) & 0.25 & 0.34 & 0.22 & 0.65
			\\ \hline
		\end{tabular}
	\end{center}
	\label{tab:m1}
\end{table}%

The second category of target imperfections which are important for these experiments are defects or engineering features which can be a seed for instabilities, which in turn inject ablator material into the hot spot. Engineering features include the fill tube\cite{doi:10.1063/1.5037816} and capsule support tent\cite{doi:10.1063/5.0017931}, which are constant for all of these experiments. Randomly varying defects include `particles', `pits', `voids', and `high-$Z$ inclusions', which are discussed here.

`Particles' are usually flakes of high-$Z$ material, predominantly from the hohlraum wall, which delaminate and fall onto the outer surface of the capsule during fabrication.\cite{Pak2020} The particle is then a seed for ablation-front instability growth, which can then inject high-$Z$ material into the fuel, which causes additional radiation losses. Particles are measured after the target is completed with x-ray imaging through the LEH with target displacements to identify, via parallax, whether a particular particle is on the capsule surface. The number and area of particles are thus characterized. An example image for the target used on N210207 is shown in Fig. \ref{fig:sagometer}. In this case there are two particles detected which are circled and labeled. In red, particle \#1 is on the capsule and has an area of 79$\mu$m$^2$. The second particle, \#2, is on the lower LEH. The particles for these four shots are summarized in Table \ref{tab:sagometer}; shots N201101 and N210220 had no particles while N210207 had the one mentioned earlier and N201122 had three. In general particles with areas $< 100$ $\mu$m$^2$ are considered relatively small, but their impact on radiative loss is not entirely understood and may be contributing to the inferred MF (e.g. in Table \ref{tab:data}).

\begin{table}[t]
	\caption{Measured high-$Z$ particles on each target shot, with areas in $\mu$m$^2$ quoted.}
	\begin{center}
		\small
		\def\arraystretch{1.2}
		\begin{tabular}{|c|c|c|c|}
			N201101 & N201122 & N210207 & N210220 \\ \hline
			$-$ & $85,55,55$ & $79$ & $-$ \\ \hline
		\end{tabular}
	\end{center}
	\label{tab:sagometer}
\end{table}
% As mentioned, instabilities due primarily to imperfections in the high density carbon capsule such as pits and voids\cite{Zylstra2020}, engineering features \cite{Pak2020,ISI:000519903500009, doi:10.1063/5.0017931, PhysRevE.95.031204, doi:10.1063/1.5032121} and particulates on the surface of the capsule lead to instability growth and the subsequent injection of ablator mass into the hot spot\cite{1748-0221-12-06-C06001, Smalyuk_2019}.

There are several categories of manufacturing defect in the capsules themselves. First, `pits' are defects of missing material on the outer surface of the shell, equivalent to a divot. As soon as the capsule is driven, these are unstable to the ablative Rayleigh-Taylor instability. `Voids' are regions of missing material within the bulk of the shell. A void can cause a seed for instability growth when the ablation front reaches it. Previous data have demonstrated that both are deleterious\cite{Zylstra2020,Zylstra2021} for implosion performance. In Fig. \ref{fig:pitsvoids} we compare the pits and voids on the actual capsules used on these experiments to batches typical of previous experiments, such as the HDC\cite{PhysRevLett.120.245003} and BigFoot\cite{Baker2018,Baker2020} (BF), which had good-quality capsules, as well as the shots reported in Ref. \citenum{Zylstra2021} (Zy21), which were marginal quality. In previous work we had to rely on batch-averaged quantities due to average metrology methodology, here we have full-sphere measurements of the pit quantities and for the larger voids. Pits are now measured with a LynceeTec digital microscopy system, voids are measured using a x-ray computed tomography system. Larger voids are measured across the full sphere with a $4 \times$ magnification tomography while the smaller voids ($<10$ $\mu$m$^3$) are taken from a $20 \times$ magnification tomography of $\sim 5\%$ of the shell.

We see that the smaller pits are reasonably comparable to capsules used in several previous works. Larger pits are not as good as the good-quality batches but are significantly improved from the marginal capsules used in Ref. \citenum{Zylstra2021}. Similar for voids, the smallest voids are comparable to previous batches while the larger voids are nearly non-existent on these shells, like previous good-quality shells. Our general understanding is that it is the larger defects which inject the most mix material and the smaller defects, while numerous, are typically below a volumetric threshold at which they could inject material.

\begin{figure}
	\centering
	\includegraphics[width=3in]{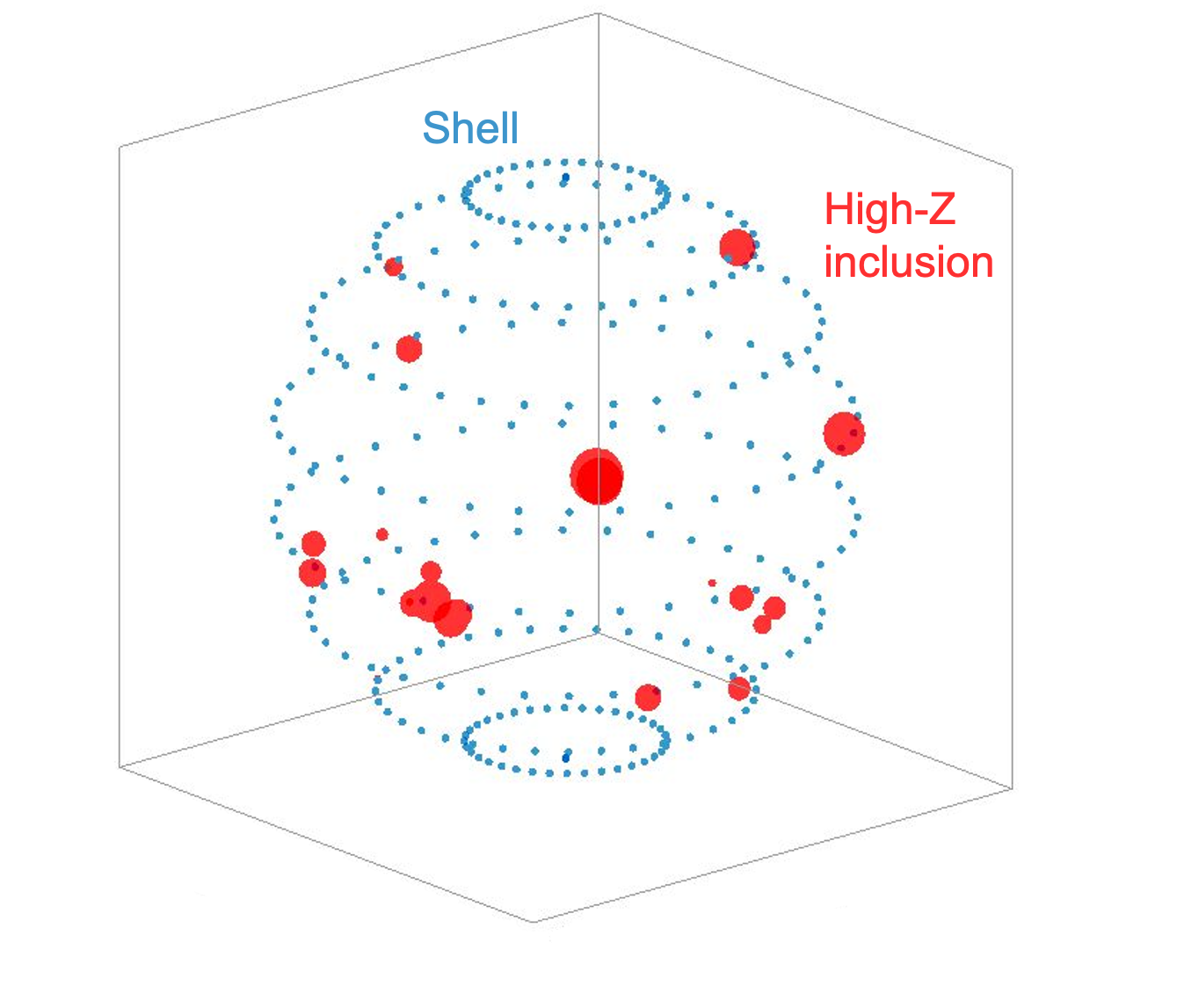}
	\caption{\label{fig:whitespots} Tomography-detected high-$Z$ inclusions on the shell used for N210207.}
\end{figure}

During this experimental series an unusual defect was detected for the capsule batch used on N210207, which included high-$Z$ `inclusions' deep in the shell, predominantly within the W-doped layer of the shell. These are likely Mo inclusions from components of the coating system that were improperly cleaned during that particular coating run. The impact of these defects on the shot is not understood at present; for these experiments, we believe that only N210207 was potentially affected. The defects are visualized using our tomography data in Fig. \ref{fig:whitespots}.

\subsection{Cryogenic fuel layering}
Several days before the shot itself, the target is installed onto a cryogenic target position (CryoTarpos) for the fueling and layering\cite{Parham2016} process. Surrounding the hohlraum is a thermo-mechanical package connected to a cryostat, which enables exquisite control over the target temperature. Nominally equimolar DT fuel, with the actual composition measured beforehand via mass spectroscopy, flows into the capsule through a fill tube, where it is condensed into a liquid. Three orthogonal x-ray imagers measure the fuel inventory, and record the growth of a concentric spherical shell of DT ice on the inside of the ablator as the temperature is reduced below the triple point, to $\sim 18.6$ K, and the fuel layer is formed via the `beta layering' technique\cite{PhysRevLett.60.1310}. The quality of the ice layer is measured and compared to specifications. The HyE (I-Raum) uses 65 (55) $\mu$m thick ice layers. Once the ice layer is complete the target is inserted into the target chamber and aligned relative to the laser beams\cite{DiNicola2012} to a tolerance in horizontal (vertical) position of $\pm 11$ ($\pm 15$) $\mu$m and orientation angles, relative to the positioner, of $\pm 0.067^\circ$. 

\begin{figure}
	\centering
	\includegraphics[width=3.5in]{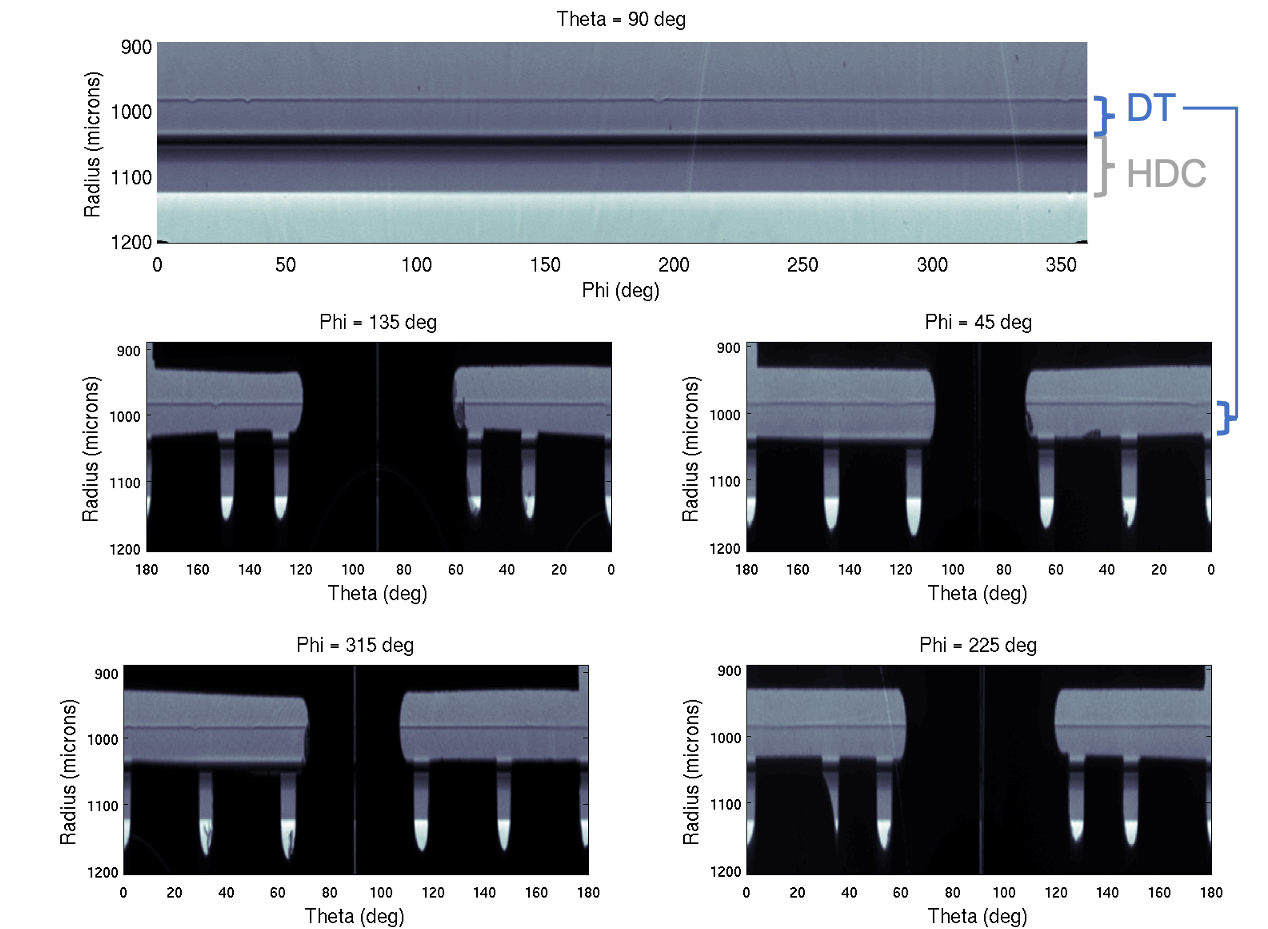}
	\caption{\label{fig:Ctps} Unwrapped images from the cryogenic layering process for N210207.}
\end{figure}

During the cryogenic layering the important quantities are first, to get the right inventory of fuel loaded, and second, to grow a high-quality concentric ice layer. Scalar quantities for the as-shot fuel composition and ice thickness are given later, in Table \ref{tab:details}. As the layer is grown, three orthogonal x-ray imaging systems record data on the low- and high-mode shape of the ice. Example images from just before shot N210207 are shown in Fig. \ref{fig:Ctps}. The images are each labeled at the top by a $\theta$ or $\phi$ value corresponding to the chamber coordinate system orientation, and are unwrapped to then be displayed as a function of the other angular coordinate. At $\theta=90^\circ$ the imaging system has a clear view of the layer through the LEH. At the equatorial plane patterns, called a `starburst', are cut out of the hohlraum wall to enable imaging along orthogonal axes. The location of the DT ice layer is marked. Feedback systems with the cryostat and localized heaters adjust temperature gradients within the target to produce a low-mode shape within spec, $\pm 0.5$ $\mu$m for modes 1-4.

\begin{figure}
	\centering
	\includegraphics[scale=1]{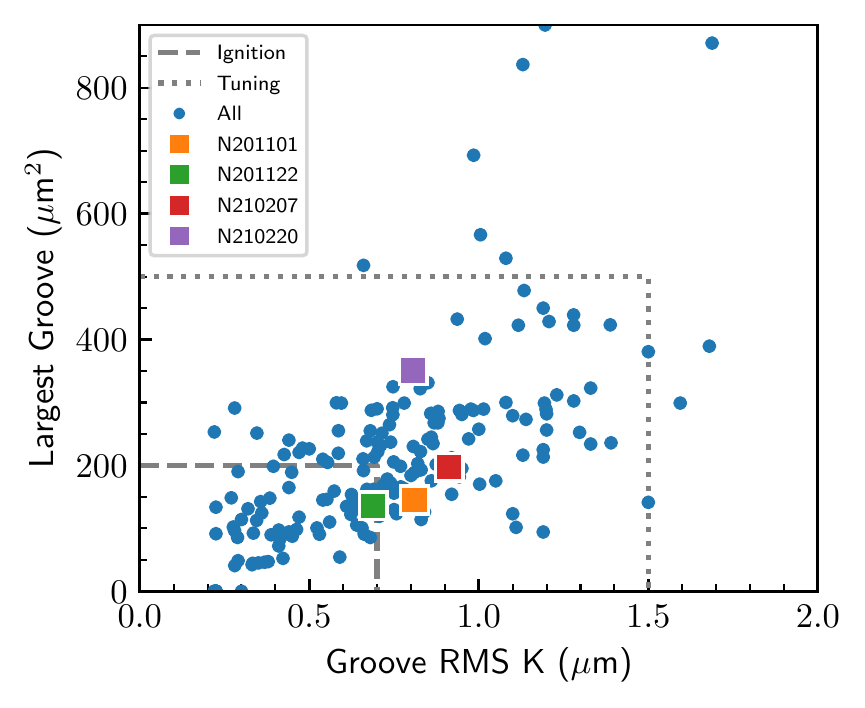}
	\caption{\label{fig:IPOM} High-mode layer quality measurements for all NIF shots and these specific experiments compared to the ignition and tuning quality specifications.}
\end{figure}

The high-mode uniformity of the layer is also important, as these features can provide seeds for hydrodynamic instabilities. In particular the layer is prone to forming grooves. A single scalar parameter $K$, with units of lenght, quantifying the RMS amplitude of these grooves was developed and published as Eq. 26 of Ref. \citenum{HaanPointDesignPOP2011} and is used, in combination with the area of the largest single defect, to characterize the high-mode layer quality. Specifications were developed for `ignition' and `tuning' specifications\cite{HaanPointDesignPOP2011}; in terms of K these are 0.7 and 1.5 $\mu$m respectively and are $200$ and $500$ $\mu$m$^2$ for the largest defect. Fig. \ref{fig:IPOM} shows the layer quality metrics relative to these specifications for these shots, as well as all prior layered experiments conducted on NIF. The final as-shot layer quality is plotted, in these experiments several attempts are typically required to obtain an adequate-quality ice layer. We can see that these experiments have comparable ice layer quality to each other, are are typical of the historical layer quality. Notably they are well within the tuning specification, while being just shy of the ignition specification. 

\begin{figure*}[t]
	\centering
	\includegraphics[scale=0.9]{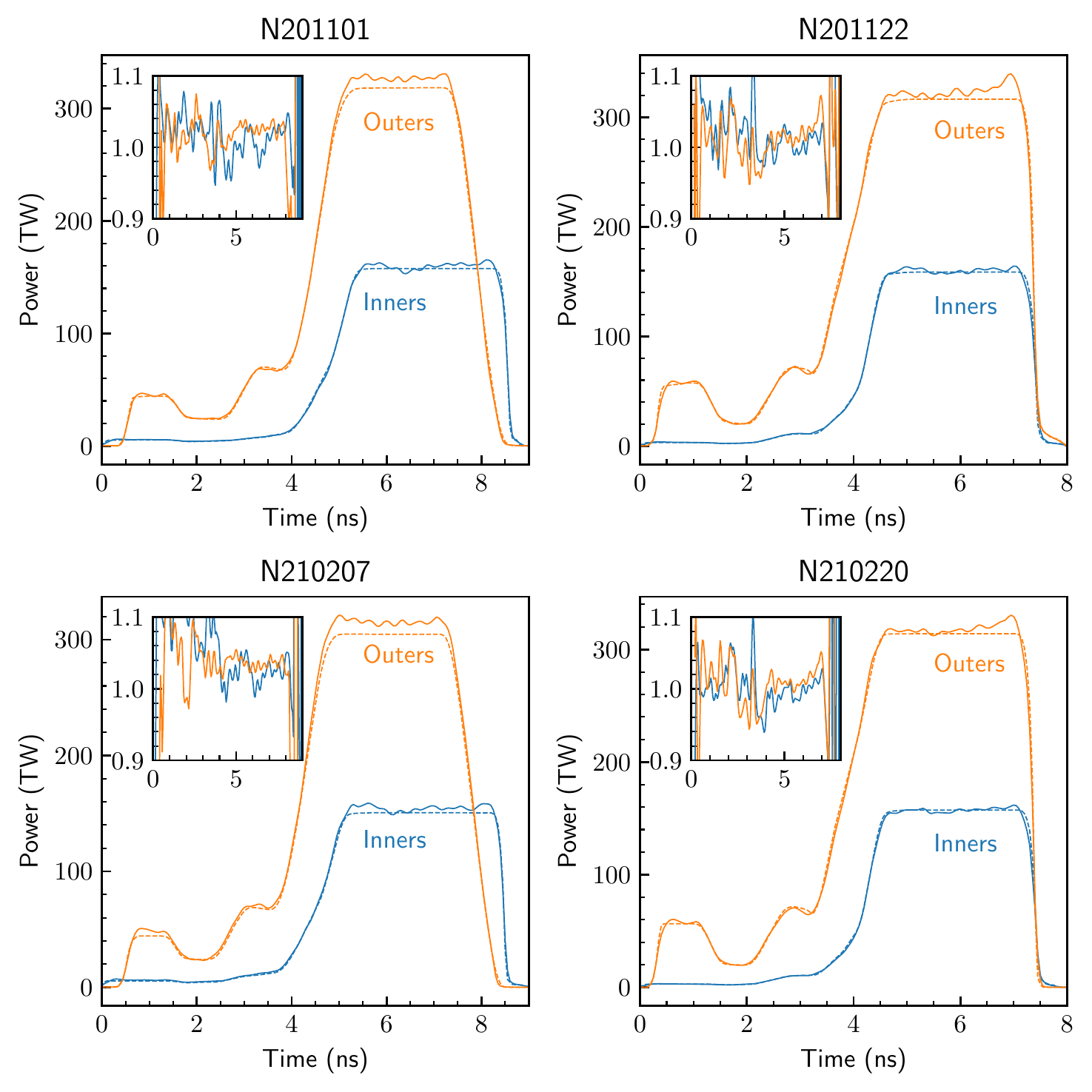}
	\caption{\label{fig:laser} Delivered laser pulses (solid lines) compared to the request (dashed lines) for inner (blue) and outer (orange) cones for all four shots. The inset axes show the ratio of the delivered to requested power as a function of time.}
\end{figure*}

\subsection{Laser delivery}
The NIF laser system\cite{doi:10.13182/FST15-144} generates the specified pulse shape, or power as a function of time, on each beam during the experiment. The pulse shape is designed to result in the desired radiation temperature history\cite{ISI:000450981300005,KritcherYoungRobeyBurningPlasmaNatPhys}. Wavelength settings, which then provide our CBET control, are controlled by the master oscillator\cite{10.1117/12.538466}. The actual laser delivery is closely monitored to ensure its sufficiently accurate compared to the request\cite{doi:10.13182/FST15-136}. As the NIF optics are damaged on high-power experiments\cite{doi:10.13182/FST15-139,doi:10.13182/FST15-119}, the lifecycle is managed to ensure the best possible performance on these high-priority experiments. The laser settings are adjusted between shots to adjust the implosion physics or compensate for changes in the target configuration using pre-shot simulations\cite{KritcherYoungRobeyBurningPlasmaNatPhys}. The laser pulse, when incident on the target, produces a peak radiation temperature of just over 300 eV. 

Example laser pulses were given, in total, in Fig. \ref{fig:setup} and are also discussed in Ref. \citenum{KritcherYoungRobeyBurningPlasmaNatPhys}. Accurate delivery of the requested laser pulse throughout its duration is key for performing high-quality experiments, as mis-delivery can affect the symmetry of the implosion - e.g. if the delivered picket energy, which launches the bubble, varies, or if the inner or outer cones deliver varying levels of energy, the $P_2$ symmetry can deviate from expectations. Inaccurate delivery can also contribute to the mode-1 asymmetry of the implosion\cite{Rinderknecht2020}. The delivery accuracy can affects the implosion parameters, especially if overall energy or power is over- or under-delivered, which affects $v_{imp}$ and $t_{coast}$. In Fig. \ref{fig:laser} we plot the actual as-delivered laser pulses for the cone averages compared to the request, plotted as actual power on the main axes and as a ratio of delivered/requested on the inset axes. In general we note that the delivery is quite good, although these experiments mostly have higher than requested outer-cone power during the peak, and N210207 experienced a noticeable picket over-delivery.

\subsection{Shot details}
Table \ref{tab:details} gives several detailed as-shot parameters for the four experiments, grouped into categories of laser, capsule, and fuel parameters. The laser parameters include the as-delivered energy and power, cone fraction (defined as the inner power during the peak divided by the total power), picket energy, pulse duration, and $\Delta \lambda$. The capsule parameters are the key dimensions of inner radius, total mass, and dopant fraction. The fuel parameters include the final layer thickness (which, with the DT density $\sim 0.25$ g/cc, determines the total fuel mass), the atomic fractions of deuterium and tritium ($f_D$ and $f_T$), and the age of the fuel. The latter is the duration since the last purge of He, which is key to avoid high levels of $^3$He accumulating in the capsule due to $\beta$ decay of tritium. 

\begin{table}[t!]
	\caption{Summary of experimental configuration parameters.}
	\begin{center}
		\small
		\begin{tabular}{|c|c|c|c|c|c|}
			\multicolumn{2}{c|}{} & N201101 & N201122 & N210207 & N210220 \\ \hline
			%\multicolumn{4}{|c|}{\textit{Laser}} \\ \hline
			\parbox[t]{3mm}{\multirow{6}{*}{\rotatebox[origin=c]{90}{\textit{Laser}}}} 
			& Energy (MJ) & 1.89 & 1.82 & 1.93 & 1.78
			\\ %\hline
			& Power (TW) & 490 & 485 & 470 & 480
			\\ %\hline
			& Cone Fraction & 32.6 & 32.9 & 32.4 & 32.8
			\\ %\hline
			& Picket (kJ) & 40.0 & 48.7 & 41.3 & 49.8
			\\ %\hline
			& Duration (ns) & 8.15 & 7.4 & 8.05 & 7.4
			\\ %\hline
			& $\Delta \lambda$ (\AA) & 1.75 & 0.5 & 1.55 & 0.5
			\\ \hline
			
			%\multicolumn{4}{|c|}{\textit{Capsule}} \\ \hline
			\parbox[t]{3mm}{\multirow{3}{*}{\rotatebox[origin=c]{90}{\textit{Capsule}}}} 
			& IR ($\mu$m) & 1049.2 & 999.9 & 1048.8 & 1000.0
			\\ %\hline
			& Mass ($\mu$g) & 3905.7 & 3738.6 & 3881.5 & 3739.9
			\\ %\hline
			& W (\%) & 0.44 & 0.42 & 0.28 & 0.42
			\\ \hline
			\parbox[t]{3mm}{\multirow{4}{*}{\rotatebox[origin=c]{90}{\textit{Fuel}}}} 
			& DT Layer ($\mu$m) & 64.7 & 55.6  & 64.2 & 55.3
			\\ %\hline
			& $f_D$ (\%) & 50.38 & 50.60 & 50.83 & 50.91
			\\ %\hline
			& $f_T$ (\%) & 49.41 & 49.19 & 48.96 & 48.88
			\\ %\hline
			& Age (hr) & 128.9 & 175.8 & 163.5 & 126.1 \\ \hline
		\end{tabular}
		\label{tab:details}
	\end{center}
\end{table}

\end{document}